\documentclass[onecolumn]{emulateapj}
\usepackage{amsmath}
\usepackage{amsbsy}
\usepackage{mathrsfs}
\shortauthors{Allred, Hawley, Abbett, Carlsson}
\shorttitle{RHD Models of Emission from M Dwarf Flares}

\newcommand{\ha}{H$\alpha$}
\newcommand{\hb}{H$\beta$}
\newcommand{\hg}{H$\gamma$}

\newcommand{\caiik}{Ca~\textsc{ii}~K}

\newcommand{\heiilya}{\ion{He}{2}~304}
\begin{document}

\title{Radiative Hydrodynamic Models of Optical and Ultraviolet
Emission from M Dwarf Flares}
\author{Joel C. Allred\altaffilmark{1,2}, Suzanne L.
Hawley\altaffilmark{3}, William P. Abbett\altaffilmark{4}, and
Mats Carlsson\altaffilmark{5}}
\altaffiltext{1}{University of Washington, Department of
   Physics, Box 351560, Seattle, WA 98195.
email: jca32@drexel.edu}
\altaffiltext{2}{Physics Department, Drexel University,
Philadelphia, PA 19104}
\altaffiltext{3}{University of Washington, Department of
   Astronomy, Box 351580, Seattle, WA 98195.}
\altaffiltext{4}{Space Sciences Laboratory, University of
California,
   Berkeley, California, 94720}
\altaffiltext{5}{Institute of Theoretical Astrophysics,
University of Oslo, P.O. Box 1029, Blindern, N-0315 Oslo,
Norway}

\begin{abstract}
We report on radiative hydrodynamic simulations of M dwarf
stellar flares and compare the model predictions to observations
of several flares. The flares were simulated by calculating the
hydrodynamic response of a model M dwarf atmosphere to a beam of
non-thermal electrons.  Radiative backwarming through
numerous soft X-ray, extreme ultraviolet, and ultraviolet
transitions are also included. The equations of
radiative transfer and statistical equilibrium are treated in
non-LTE for many transitions of hydrogen, helium and the
\ion{Ca}{2} ion allowing the calculation of detailed line profiles and
continuum radiation.  Two simulations were carried out, with
electron beam fluxes corresponding to moderate and strong beam
heating.  In both cases we find the dynamics can be naturally
divided into two phases: an initial gentle phase in which
hydrogen and helium radiate away much of the beam
energy, and an explosive phase characterized by large
hydrodynamic waves.  During the initial phase, lower
chromospheric material is evaporated into higher regions of the
atmosphere causing many lines and continua to brighten
dramatically.  The \heiilya\ line is especially enhanced, becoming
the brightest line in the flaring spectrum.  The hydrogen Balmer
lines also become much brighter and show very broad line widths,
in agreement with observations.  We compare our predicted
Balmer decrements to decrements calculated for several
flare observations and find the predictions to be in general
agreement with the observations. During the explosive phase
both condensation and evaporation waves are produced.  The
moderate flare simulation predicts a peak evaporation wave of
$\sim$130 km s$^{-1}$ and a condensation wave of $\sim$30 km
s$^{-1}$. The velocity of the condensation wave matches
velocities observed in several transition region lines. The
optical continuum also greatly intensifies, reaching a peak
increase of 130\% (at 6000 \AA) for the strong flare, but does
not match observed white light spectra.
\end{abstract}

\section{Introduction}
Solar-type flares are common on cool stars with outer convective
envelopes.  Extensive observations of flares on magnetically
active M dwarfs (dMe) in the solar neighborhood have
revealed many similarities including the presence of
enhanced and broadened Balmer emission, increases in optical
white light and radio emission, and similar time scales for the
impulsive and gradual phases. A detailed comparison of solar and
stellar flares is presented in \citet{1991ARA&A..29..275H}. The
generally accepted model of solar flares assumes that
reconnecting magnetic fields cause the acceleration of electrons
(and possibly protons).  The accelerated electrons travel down
magnetic field lines, impacting in the lower atmosphere, producing
hard X-ray bremsstrahlung and causing increased optical
continuum and line emission.  As the electrons heat the lower
atmosphere, it expands and ``evaporates'' into the
transition region and corona, thereby increasing the thermal soft
X-ray flux.  The Neupert effect describes the relationship
between hard and soft X-rays on the Sun \citep{1968ApJ...153L..59N,
1999ApJ...514..472M}, and stellar observations of the Neupert
effect \citep{h95, g96, ha03, 2005A&A...431..679M}
provide evidence that stellar and solar flare processes show
similar effects.

Even though solar and stellar flares are likely to be produced by
the same basic mechanisms, the different physical conditions
on M dwarfs compared to the Sun give rise to a number of
observational differences.  Flares on dMe stars occur more
frequently and are often much more energetic than solar flares.
This is likely the result of the stronger magnetic fields
on dMe stars \citep{1996ApJ...459L..95J} which produce more
magnetic activity than on the Sun. For stellar flares, increases
in the optical continuum are very common; this is often taken as
the defining observational signature of a flaring star. However,
detectable elevations in the optical continuum are rare
in solar flares, since the Sun has much brighter background
continuum emission, and therefore only the largest flares produce
enough white light contrast to be seen.

On the Sun, impulsive hard X-rays are spatially and temporally
correlated with increased white light emission
\citep{1992PASJ...44L..77H}.  By extension, the rapid rise in
near ultraviolet and blue emission seen in stellar flares is
thought to correspond with the solar impulsive phase.  Therefore,
stellar
optical flares are likely to be the analog to solar white light
flares.  We recently developed models of impulsive solar white
light flares \citep[hereafter A05]{a05}, and in this paper we
apply those techniques to model impulsive M dwarf flares.  Previous
and recent M dwarf flare observations
\citetext{\citealp{1984iue..conf..247R, d88, 1988MNRAS.235..573P,
hp91, e92, 1995A&AS..114..509A, 1998A&A...338.1057J, ha03}}
(hereafter, we refer to \citealp{ha03} as H03) are used to provide
observational constraints on the models.

The paper is organized as follows.  In Section 2, we briefly
outline the method by which the radiative hydrodynamic equations
are solved. We also describe the procedure used to generate the
initial atmosphere and the methods for incorporating soft X-ray
and electron beam heating into the simulations.  In Section 3, we
discuss the dynamics of our simulations and present detailed line
and continuum profiles. In Section 4, we compare the M dwarf
models to the solar models of A05, while in Section 5 we compare
predictions of our models to stellar flare observations. In
Section 6 we present our conclusions.

\section{Method of Solution \label{sec:method}}
The method we have used is described in detail in A05 and
\citet{ah99}, and we refer the reader to those papers for a more
detailed discussion.  The equations of hydrodynamics, statistical
equilibrium and radiative transfer are solved on a one-dimensional
adaptive grid using the RADYN code \citep{cs94,cs95,cs97}.

The preflare atmosphere was generated in a manner similar to AH99.
We assume that the plasma moves along field lines in a flux tube
which we approximate as a one-dimensional cylinder.  Our model
flux tube extends from the photosphere to the corona and has a
length of $10^9$ cm. The preflare atmosphere of \citet{hf92} was
used as the initial starting state for the atmospheric structure.
Constant non-radiative quiescent heating was applied to grid zones
with photospheric column mass (i.e. column mass greater than
3.16~g~cm$^{-2}$) to balance the energy losses in the photosphere.
The upper boundary was held at $6 \times 10^6$ K. With these
boundary conditions and no external sources of heating, the
atmosphere was allowed to relax to a state of hydrostatic
equilibrium.

We modeled a flare by calculating the radiative hydrodynamic
response to a beam of non-thermal electrons injected at the top of
the loop.  The electron beam heats the stellar atmosphere which
increases the soft X-ray, extreme ultraviolet and ultraviolet
radiation (hereafter, we refer to these collectively as XEUV flux);
this further heats the atmosphere through radiative backwarming.
For the dynamic simulations, we change the upper boundary
condition to a transmitting boundary and remove its
temperature constraint. We note that with no external source of
coronal heating our coronal model will cool.  However, the time
scale for coronal cooling is much greater than the impulsive
time scale \citep{1994ApJ...422..381C}. As the flare evolves,
material flows through the upper boundary, emulating the expansion
of the flux tube. Since our flux tube has a fixed length of
$10^9$~cm, the loop apex and the upper corona move outside the
region of computation during the course of the flare simulation. 
Our concern in this paper is modeling optical and UV emission
which primarily originate from the lower atmosphere. In
\citet{a05}, we found that chromospheric heating from the upper
corona was negligible when compared to the energy deposited by the
electron beam.  Thus, the ``loss'' of the upper corona is not
likely to affect the chromospheric emission. 

The non-LTE atomic level populations are obtained for a six level
with continuum hydrogen atom, nine level with continuum helium
atom and six level with continuum \ion{Ca}{2} ion using the
technique of \citet{sc85}. We calculate numerous bound-bound and
bound-free transitions in detail.  These are listed in
Tables~1 and 2 of A05. Complete redistribution is assumed for
all transitions, but the effects of partial redistribution are
approximated for the Lyman transitions by truncating the line
profiles at 10 Doppler widths. Other transitions are included as
background opacity in LTE.  The opacities were obtained from the
Uppsala package of \citet{g73}. We include optically-thin
radiative cooling from bremsstrahlung and coronal metal
transitions using emissivities acquired from the ATOMDB database
\citep{sblr01}. The equations are solved on an adaptive grid using
250 grid points in depth, 5 in angle and up to 100 in frequency
for each transition.

Numerous observations of hard X-rays during \emph{solar} flares
indicate the presence of downward directed non-thermal electrons
impacting in the lower solar atmosphere.  Unfortunately, current
instruments are unable to detect hard X-rays produced from flares
on stars other than the Sun. However, as mentioned above,
observations of the Neupert effect provide justification that the
heating mechanisms may be the same. Therefore, in these models we
employ an electron beam as the source of heating in the lower
stellar atmosphere.  In particular, we use an electron beam with
energy spectrum obtained by \citet{h03} for the 23 July 2002
X-class \emph{solar} flare observed by RHESSI.
The injected electrons are found to
have a double power law energy distribution of the form
\begin{equation}
\label{eqn:dpl}
 F_0(E_0) = \frac{ \mathcal{F} (\delta_u-2) (\delta_l-2)}
{E_c^2 \left( (\delta_u-2) - \left( \frac{E_B}{E_c}
\right)^{2-\delta_l} (\delta_u-\delta_l) \right) } \left\{
\begin{array}{ll}
\left( \frac{E_0}{E_c} \right) ^{-\delta_l} & \textrm{for
$E_0<E_B$} \\\left(\frac{E_B}{E_c} \right )^{\delta_u-\delta_l}
\left(\frac{E_0}{E_c} \right) ^{-\delta_u} & \textrm{for
$E_0>E_B$}\end{array} \right.
\end{equation}
where $\mathcal{F}$ is the electron energy flux that enters the
magnetic loop; $E_c$ is the cutoff energy below which the X-ray
emission is assumed to be thermal; and $E_B$ is the break
energy where the distribution shifts from spectral index
$\delta_l$ to $\delta_u$.  We take $\delta_l$, $\delta_u$, $E_c$,
$E_B$ to be constant with values of 3.0, 4.0, 37 keV and 105 keV
respectively corresponding to the peak of the 23 July 2002
flare.  The rate of energy deposited by the electron beam is
modeled using the technique of \citet{e78}.

During flares, the density in the transition region and corona
becomes elevated increasing the amount of XEUV photons originating
from these regions. The outward-directed photons we detect as
enhanced emission, while the downward-directed photons heat the
lower atmosphere through increased photoionizations. We have used
the ATOMDB database to
determine the thermal volume monochromatic emissivity as a
function
of wavelength and temperature. We include line emissivities
calculated for approximately 34,000 transitions at 37 temperature
points ranging from $10^4$ K to $10^7$ K.  We have grouped the
transitions into 14 wavelength bins with a range of 1 -- 2500
\AA. Table~3 in A05 lists the range and emissivity-weighted
central wavelengths of each bin. The XEUV backwarming rate is
calculated from Equation~6 of A05 with the assumption that the
energy from the photoionized electrons is entirely converted to
heat.

\section{Flare Simulations}
We carried out two flare simulations, corresponding to moderate
beam heating with an electron flux, $\mathcal{F}=10^{10}$ ergs
cm$^{-2}$ s$^{-1}$ (the F10 flare) and strong beam heating with
$\mathcal{F}= 10^{11}$ ergs cm$^{-2}$ s$^{-1}$ (the F11 flare).
The F10 flare was evolved for 230 s and the F11 flare for 15 s.
In both cases the beam energy spectrum is given by the double
power law as described in Equation~\ref{eqn:dpl}.

\subsection{F10 Flare Dynamics}
\begin{figure}
\plotone{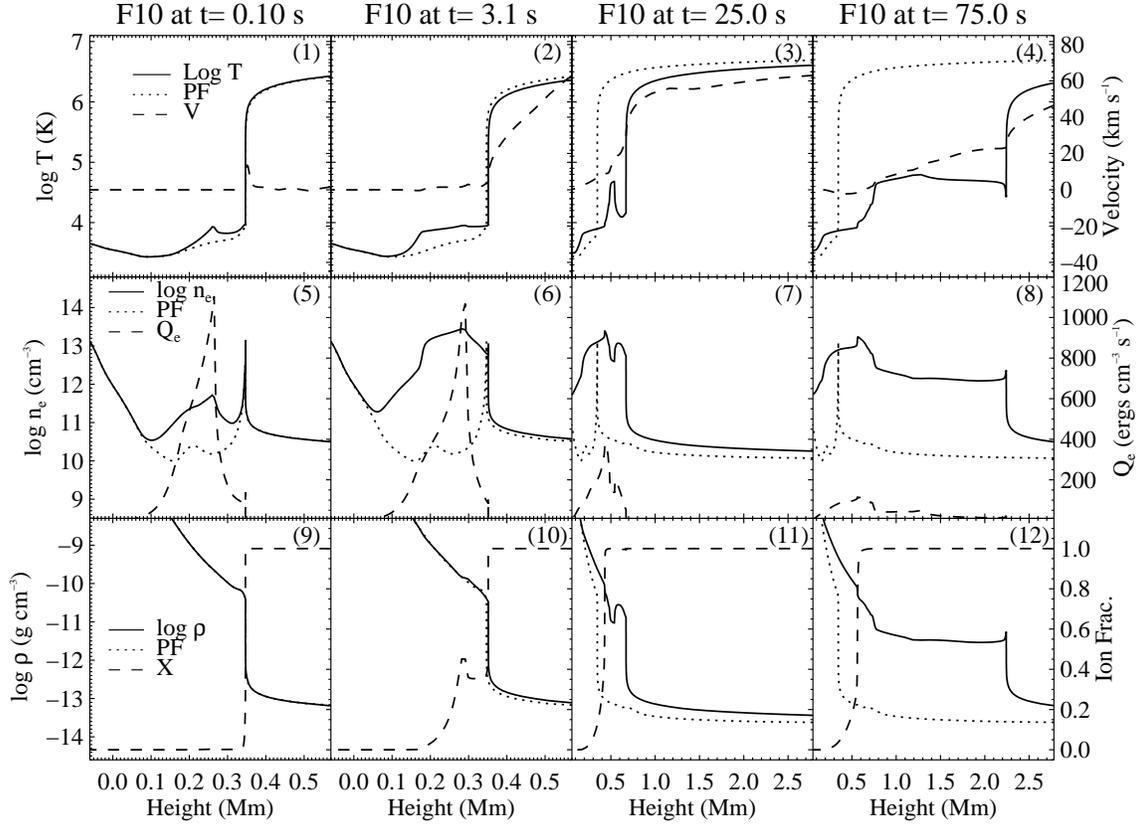}
\caption{The stellar atmosphere at four times during the
initial phase of the F10 flare.  The top row shows the log of the
temperature, $T$ (left axis) and velocity (right axis) as a
function of height compared with the preflare state (PF).  In the
middle row the electron density, $n_e$ (left axis) and beam
heating rate, $Q_e$ (right axis) are
plotted.  The bottom row shows the mass density, $\rho$ (left
axis) and hydrogen ionization fraction, $X$, (right axis). Note
the change in scale of the horizontal axis in the third and
fourth columns.}
\label{fig:f10gen}
\end{figure}

Figure~\ref{fig:f10gen} shows the evolution of the temperature
and density stratification, the electron beam energy deposition
rate, the ionization fraction, and the electron density during the
F10 flare.  The electron beam initially penetrates to a depth of
0.26 Mm, and the surrounding temperature quickly rises in response
(the first column of Fig.~\ref{fig:f10gen}). A large hydrodynamic
wave forms and begins to carry material from the chromosphere into
the transition region.  By 0.25 s, the temperature plateaus at
$\sim$9000 K in the region of beam energy deposition as a result of
elevated hydrogen radiative cooling. For the next several seconds,
the flare evolves slowly as much of the beam energy is radiated
away by hydrogen emission (the second column of
Fig.~\ref{fig:f10gen}). This results in significantly stronger
hydrogen emission lines. The transition region continues to fill
up with plasma evaporated from the chromosphere and becomes
extended in height (see panels 3 and 4 of Fig.~\ref{fig:f10gen}).
By 10.0 s most of the hydrogen in the region of beam heating is
ionized, and the radiative cooling is no longer able to balance the
beam heating.  The atmosphere rapidly heats to a temperature of
$\sim50,000$ K, but by 25.0 s it again plateaus (third column of
Fig.~\ref{fig:f10gen}). This time the equilibrium is a result of
helium radiative cooling, and a relatively gentle phase begins as
hydrogen and helium radiate away most of the beam energy. We note
that although the plasma in the extended transition region
(z=0.8 -- 2.2 Mm in panel 4 of Fig.~\ref{fig:f10gen}) appears to
have ``cooled,'' when compared to the preflare atmosphere, this is
not the case. Material from lower, cooler regions is brought
upward and heated. For example, the plasma located in this region
has a temperature of about 60,000 K at 75.0~s, but originates in
the chromosphere from a height of 0.3 Mm and a temperature of
5000~K. In general, when comparing the atmospheric structure
during the flare evolution,
care must be taken to consider that the plasma has moved.

\begin{figure}
\plotone{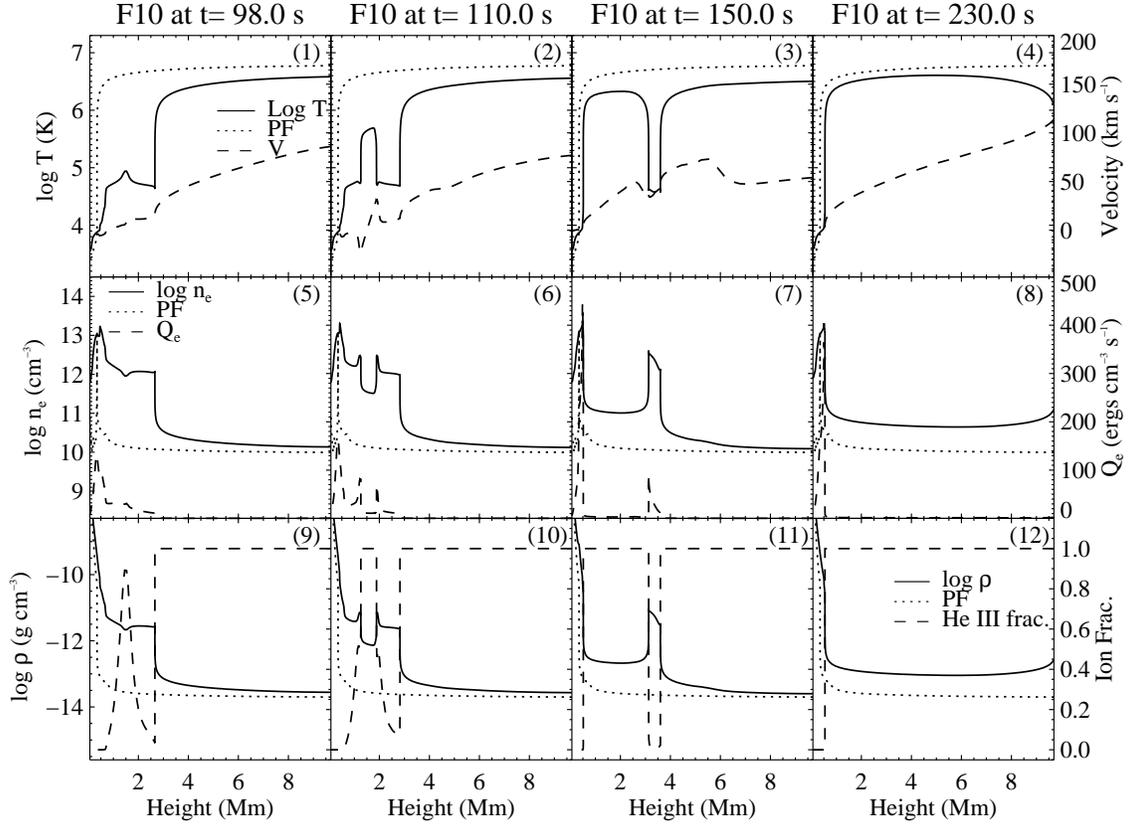}
\caption{The F10 flare atmosphere at four times during the
explosive phase.  The quantities plotted are identical to
Fig.~\ref{fig:f10gen} except that the \ion{He}{3} fraction is
plotted in the bottom row rather the hydrogen ionization
fraction.}
\label{fig:f10exp}
\end{figure}

By 98~s, the helium in the region of beam heating has become
almost completely ionized and can no longer effectively cool the
atmosphere (panel 9 of Fig.~\ref{fig:f10exp}).  The atmosphere
responds by explosively heating. The temperature quickly rises
(panel 2 of Fig.~\ref{fig:f10exp}), and a large hydrodynamic wave
is created. We refer to this period as the explosive phase.
Explosive phase dynamics are shown in Figure~\ref{fig:f10exp}. A
high temperature bubble forms in the region near 1.5 Mm
(second column of Fig.~\ref{fig:f10exp}) and for the next 130 s
the bubble expands as the wave propagates through the atmosphere.
This stellar flare simulation was not able to track
the wave passing through the top of flux tube, and did not
attain a steady state.  Our final atmosphere (last column of
Fig.~\ref{fig:f10exp}) has a denser corona and transition region
(panel 8) than the preflare atmosphere. We note that much of the
original corona has flowed through the original flux tube boundary.
In A05 we found, for solar flare simulations, that after the wave
passed through the boundary of the flux tube the loop heated
quickly.  The solar flare loop top eventually reached a steady state
temperature approximately five times the preflare value, and we
expect a similar result for stellar flare loops.

\begin{figure}
\plotone{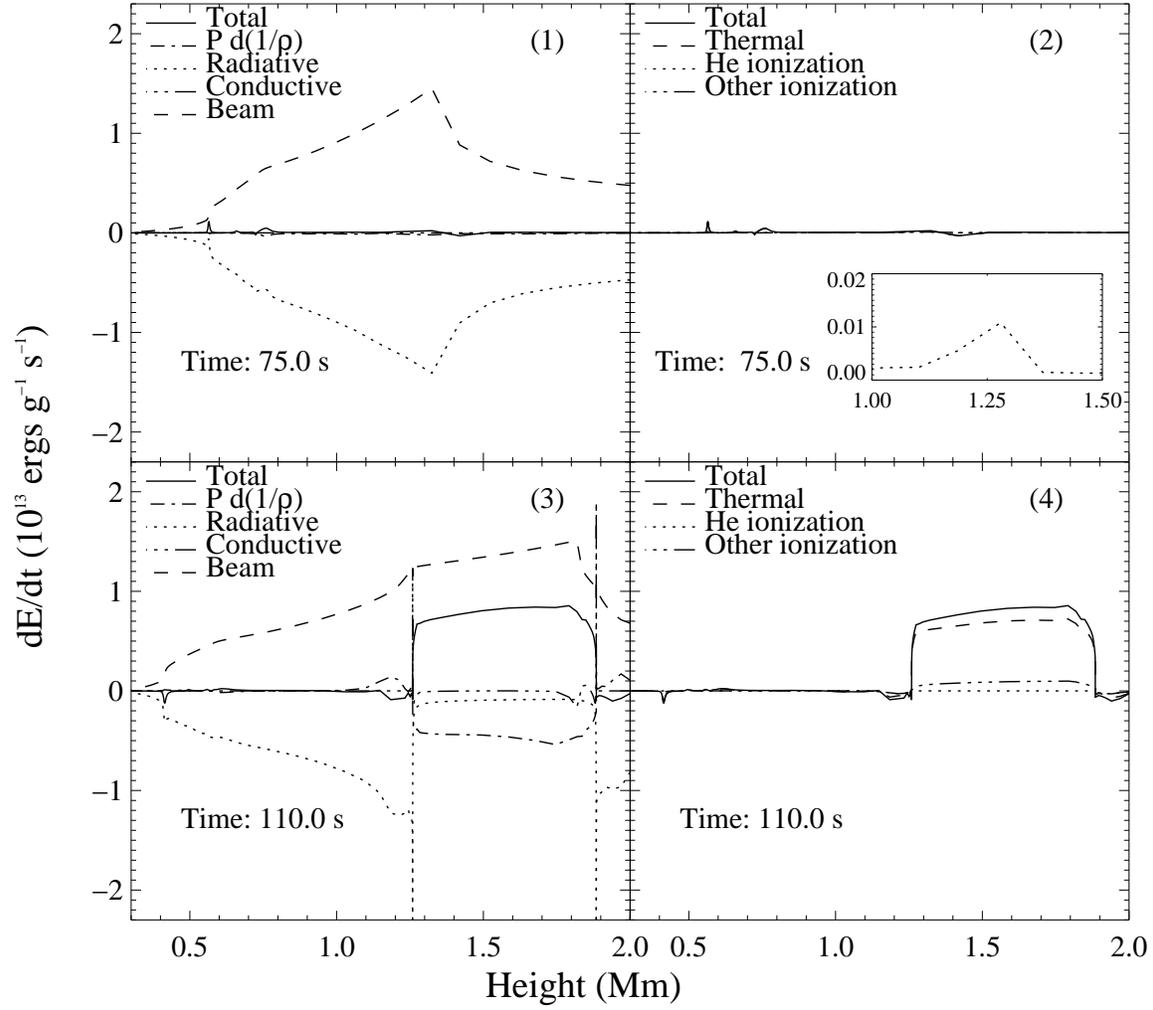}
\caption{Energetics for the F10 flare during the gentle phase
(panels [1,2]) and the explosive phase (panels [3,4]).  The
atmospheric structure for panels [1,2] is shown in the
last column of Figure~\ref{fig:f10gen}, and the structure
for panels [3,4] is shown in the second column of
Figure~\ref{fig:f10exp}.  Important terms in the energy
conservation equation are shown in panels [1,3], and contributions
to the change in internal energy are shown in panels [2,4].  The
inset in panel [2] shows a closeup view of the rise in internal
energy due to increased He ionization.}
\label{fig:ebal}
\end{figure}

The gentle and explosive phase energetics are plotted in
Figure~\ref{fig:ebal}.  During the gentle phase, the energy
deposited by the beam is essentially balanced by radiative
cooling (panel 1). A small amount of energy goes into increasing
the helium ionization in the region of 1.0 -- 1.5 Mm (see the inset
in panel 2). Once the helium is completely ionized, the radiative
cooling sharply drops, and the explosive phase begins. In the
explosive phase, there is a significant portion of the beam energy
in the region of the expanding high temperature bubble (see panel 2
of Fig.~\ref{fig:f10exp}) that is not radiated away (panel 3 of
Fig.~\ref{fig:ebal}). Most of this energy goes into increasing the
temperature (panel 4) and expanding the bubble (see P d(1/$\rho$)
in panel 3).

\subsection{F11 Flare Dynamics}
\begin{figure}
\plotone{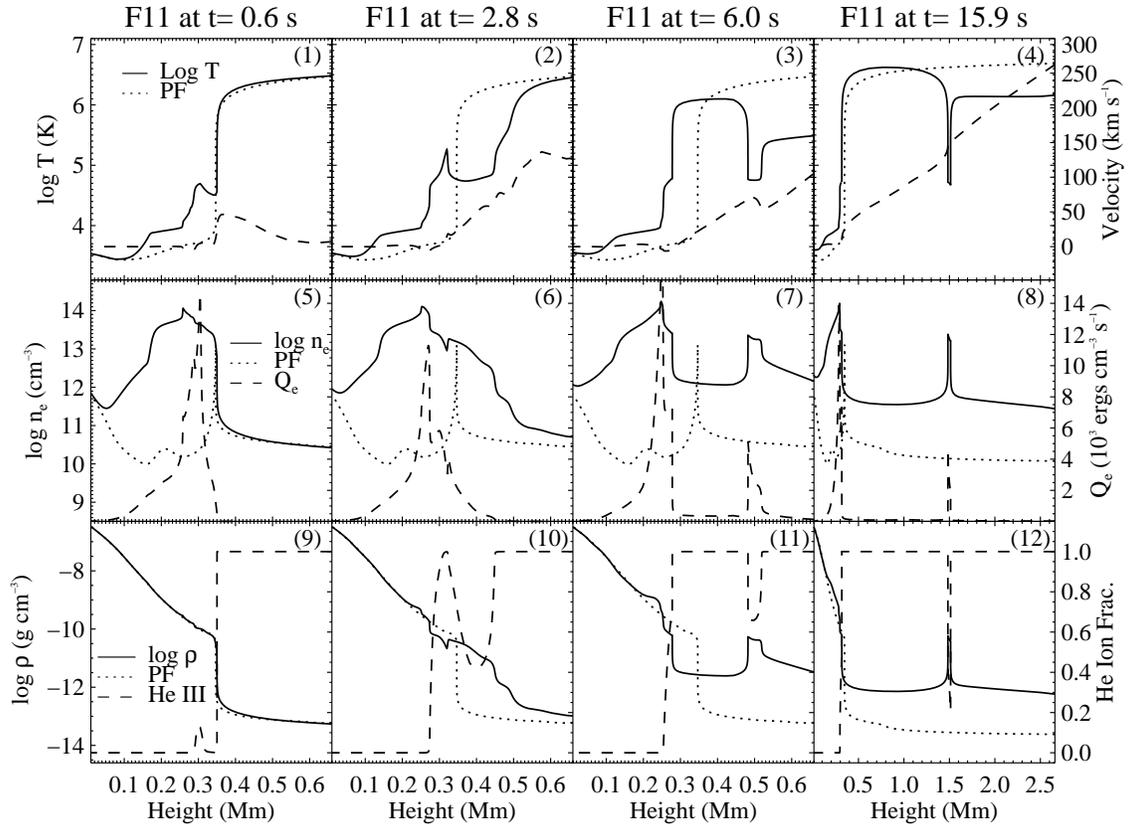}
\caption{The flare atmosphere at four times during F11.   The
quantities plotted are identical to Fig.~\ref{fig:f10gen} except
that the \ion{He}{3} fraction is plotted in the bottom row rather
the hydrogen ionization fraction.}
\label{fig:f11atm}
\end{figure}

The evolution of the F11 flare is similar to F10, but proceeds much
faster.  Figure~\ref{fig:f11atm} shows the atmosphere at several
times during this flare.  As in F10, the beam initially penetrates
to a height of 0.26 Mm, and in response the atmosphere rapidly heats
to about 10,000 K.  A strong hydrodynamic wave begins to carry
material into the transition region. By 0.03~s, the temperature
increase has stabilized as the beam heating is in equilibrium with
hydrogen radiative cooling. This equilibrium lasts until 0.16 s when
most of the hydrogen in the region of beam heating has been ionized.
In a fashion similar to F10, the atmosphere again quickly heats but
by 0.5 s plateaus at 50,000 K as a result of an equilibrium of beam
heating with helium radiative cooling (first column of
Fig.~\ref{fig:f11atm}). By 2.5 s, helium in the region of 0.3 Mm is
completely ionized, and the atmosphere undergoes explosive
heating (second column of Fig.~\ref{fig:f11atm}). As in F10, a
high temperature bubble forms and propagates toward the boundary of
the flux tube. A notable difference between F10 and F11 is the
height at which this happens: $\sim 2$ Mm in F10 compared to 0.3
Mm in F11, see panel 2 of Fig.~\ref{fig:f10exp} and panel 2 of
Fig.~\ref{fig:f11atm}. The evolution of F11 is much quicker than
F10, so less material has had time to evaporate into the upper
atmosphere. The beam therefore, deposits most of its energy
lower in the atmosphere in F11 than in F10.  At 15.9~s there is a
very steep decrease in temperature at $z=1.5$ Mm (panel 4). This is
caused by the high level of radiative cooling due to increased
density at the front of the outward-directed explosive wave.  This
is likely an artifact of the simulation, as strongly blue-shifted
transition region material is not seen in flare observations (see
\S~\ref{sec:uvlvel}).

\subsection{Line Profiles}
\begin{figure}
\plotone{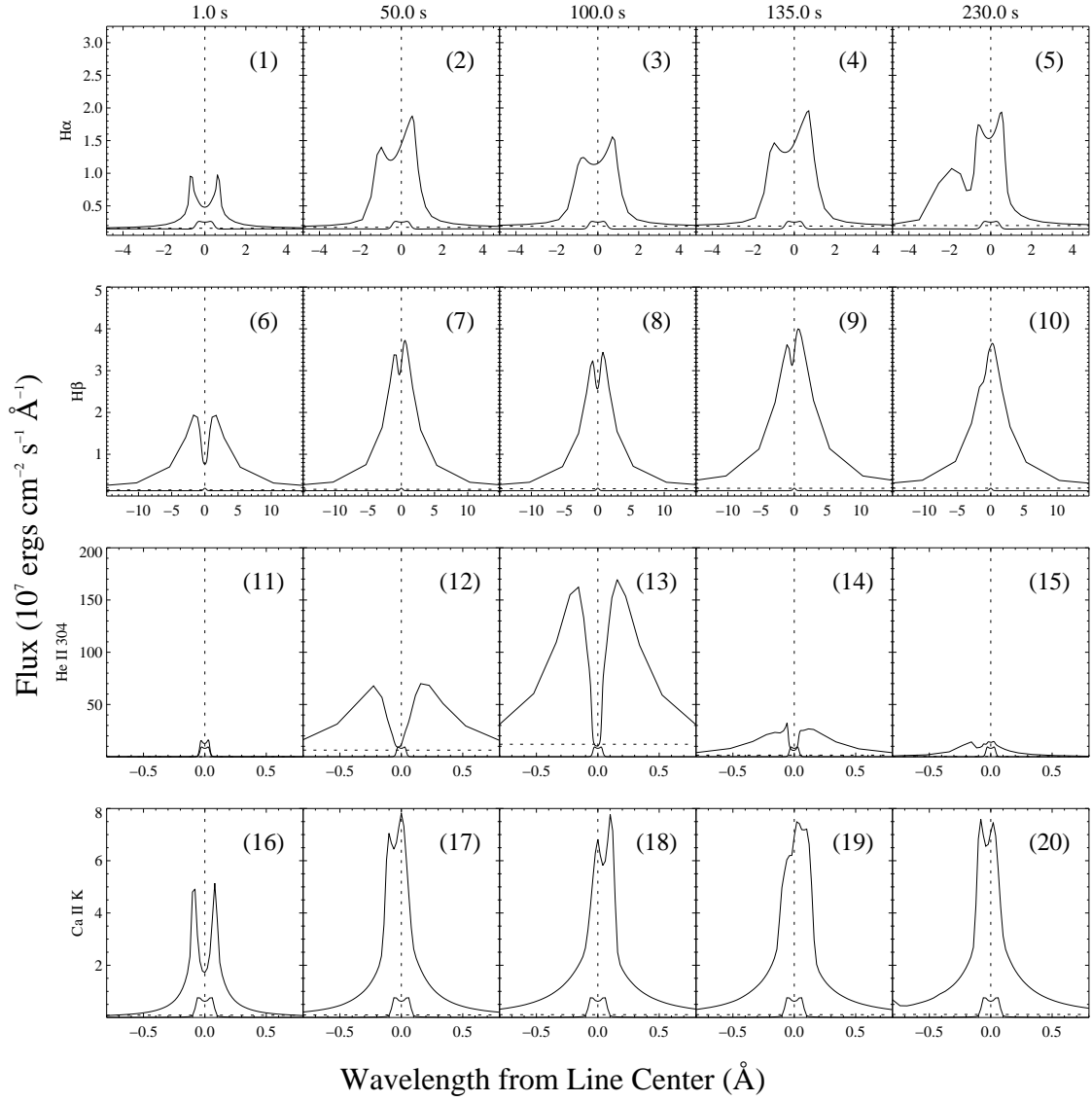}
\caption{Line profiles are plotted for \ha, \hb, \heiilya,
and \caiik\ at five times during F10.  The smaller line
profile in each panel is the quiescent line profile. The
horizontal dotted line indicates the flare continuum level, and
the vertical dotted line is line center.}
\label{fig:lprof}
\end{figure}

Mass motions during the flare simulations cause significant
Doppler shifts and introduce large asymmetries in the line
profiles. Figure~\ref{fig:lprof} shows line profiles for four
important lines during F10.  The first column (panels 1, 6, 11 and
16) show that at 1.0 s these emission lines have increased in
intensity, but there are no large mass motions since the line
profiles are relatively symmetric.  In the second column, however,
the line profiles are significantly distorted as a result of large
velocities produced by flare heating.  The line profiles are
especially complicated when they are formed in regions of
oppositely directed waves. For example, the \ha\ profile in panel
2 has an enhanced red wing indicating downward-directed plasma,
but its line center is blue-shifted. The line center is formed
higher in the stellar atmosphere than the wings, where the plasma
is ``evaporating'' into the transition region.  In panel 5, \ha\
has a very large blue wing.  This is the result of the high
velocity, high density evaporation wave passing through the
boundary of the flux tube.  \hb\ responds to flare heating very
quickly.  By 1.0 s (panel 6), it is already about 150 times
brighter than in the preflare atmosphere.  Also note that \hb\
becomes very broad, as discussed in more detail in
\S~\ref{sec:lbroad}. The \heiilya\ line becomes very bright
as a result of the extended transition region that forms during
the gentle phase (see panel 4 of Fig.~\ref{fig:f10gen}). During
this phase, it becomes the most intense line in the flaring
spectrum. The \heiilya\ flux reaches a maximum at the onset of the
explosive phase (Fig.~\ref{fig:lprof}, panel 13). Once the
explosive phase begins, the atmosphere heats beyond 50,000 K (i.e.
the temperature of
formation of \ion{He}{2}) and the intensity decreases (panels 14
and 15).  In panel 14, a 22 km~s$^{-1}$ wave produced at the onset
of the explosive phase causes an enhanced blue peak in the
\heiilya\ line profile. As this wave passes into the higher
temperature regions of the corona, it begins to move faster.  By
230 s, the wave is moving at 129 km~s$^{-1}$ and \heiilya\ shows
a strong blue-shift (panel 15).  In contrast, the \caiik\ line is
sensitive to velocities in the chromosphere. In panel 17, \caiik\
is blue-shifted indicating evaporation of chromospheric material
into the transition region. The \caiik\ line profiles in panels
18 and 19 are red-shifted as a result of chromospheric
condensation waves. At 100~s, a relatively slow
($\sim$6~km~s$^{-1}$) downward directed wave produced from the
initial beam impact passes through the region of \caiik\
formation producing the red shift (panel 18). At 135~s, a much
faster condensation wave ($\sim$30~km~s$^{-1}$) produced from the
explosive increase in temperature passes through this region and
\caiik\ is further red-shifted (panel 19).

\subsection{Continuum \label{sec:cont}}
\begin{figure}
\epsscale{1}
\plottwo{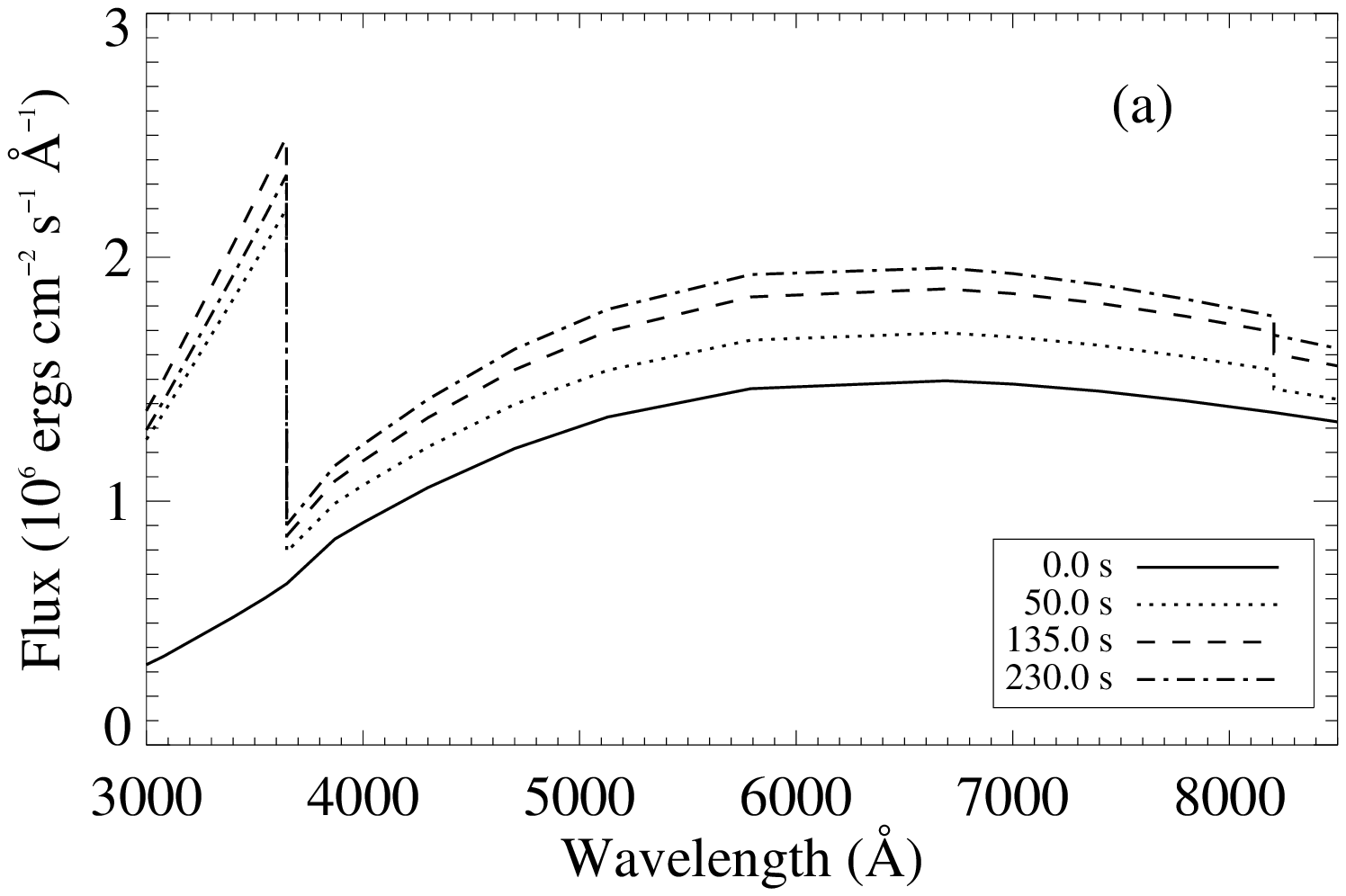}{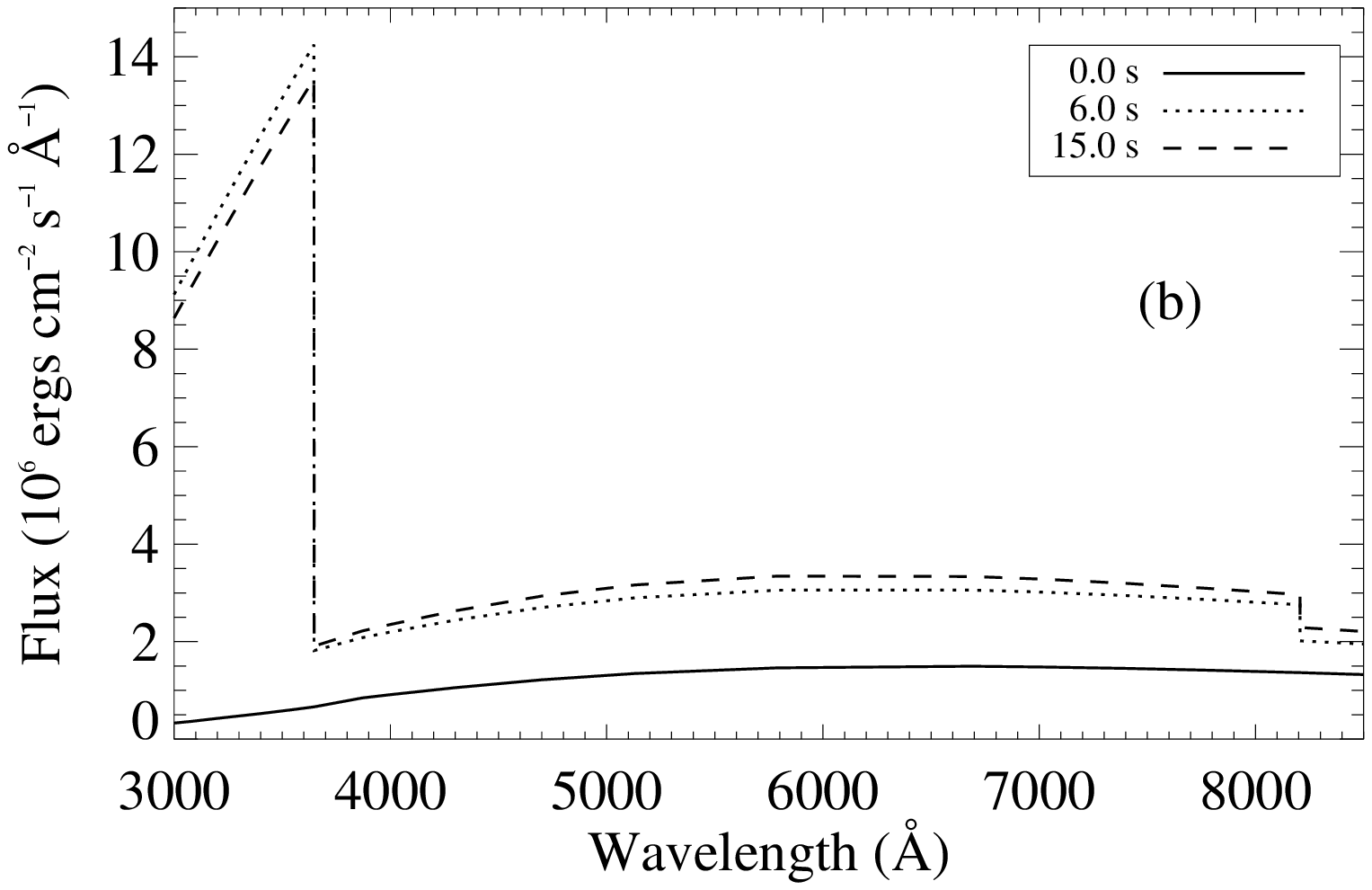}
\caption{Continuum spectra are
plotted for the preflare atmosphere and a few times during the F10
(a) and F11 (b) flares. The spectra exhibit a large Balmer jump
(3646 \AA), a noticeable Paschen jump (8205 \AA) and increased
blackbody emission.} \label{fig:cspec}
\end{figure}

\begin{figure}
\plottwo{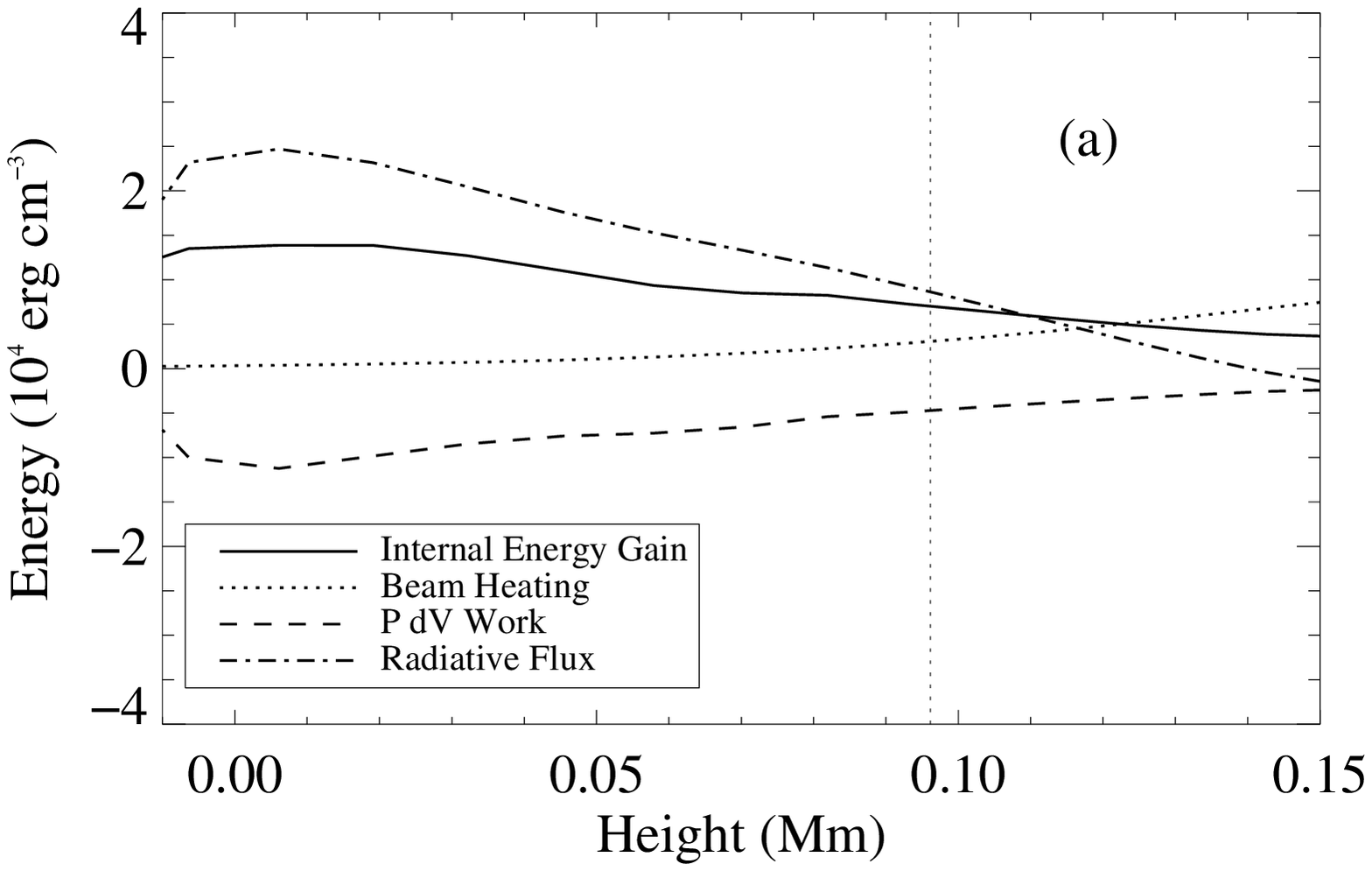}{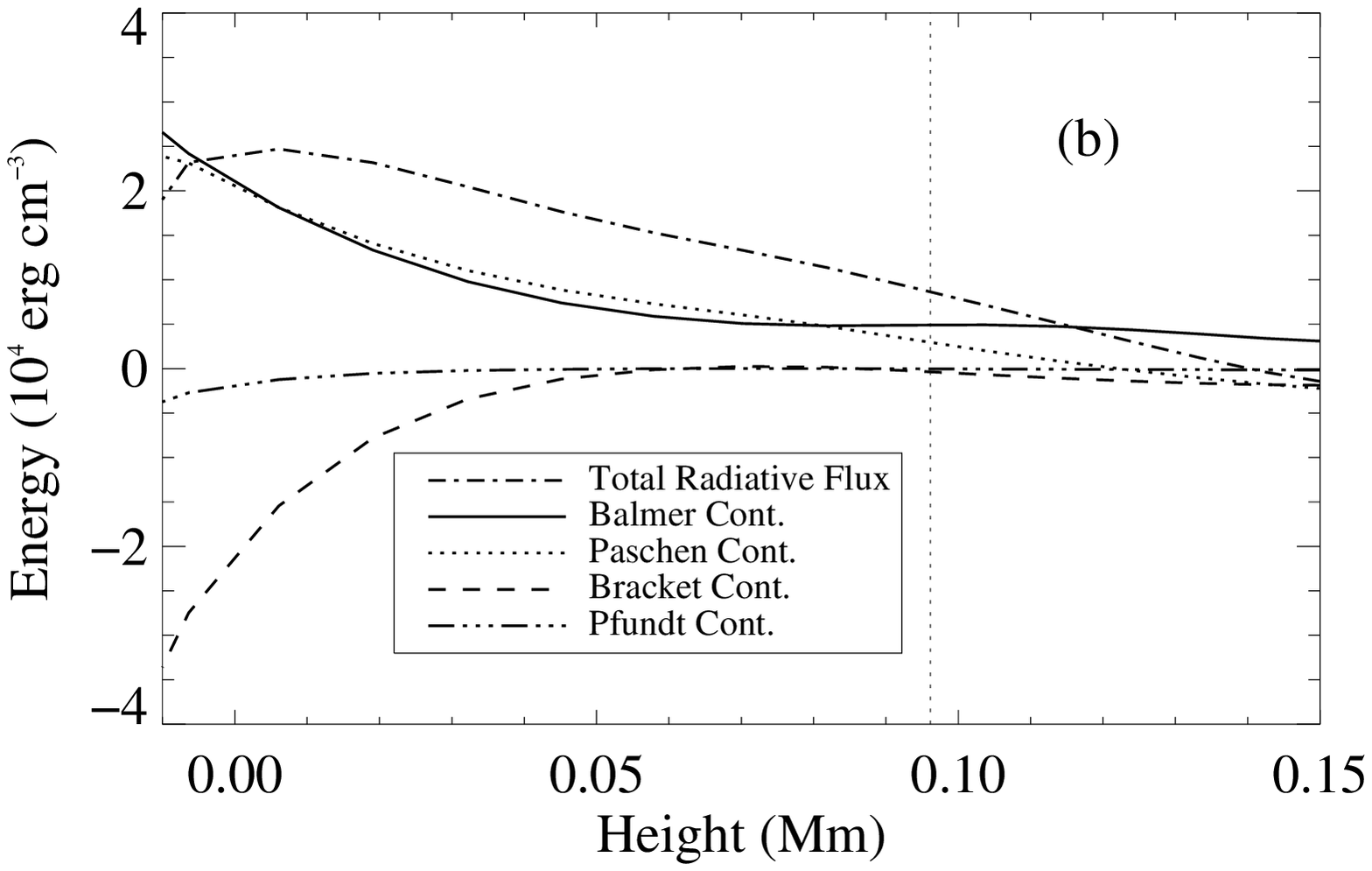}
\caption{Internal energy gain in the lower atmosphere as a function
of height above the photosphere. (a) The solid line represents the
total internal energy gain. The dotted line is the energy deposited
by the beam.  The dashed line is P dV work, and the dot-dashed line
is the radiative flux. The heating from radiation dominates in the
photosphere. (b) Contributions to the total radiative flux. The
Balmer, Paschen, Bracket and Pfundt continua are represented by the
solid, dotted, dashed, and dot-dot-dot-dashed lines respectively.
The Balmer and Paschen continua are the dominant sources of heating
in this region. Lyman continuum does not significantly contribute
to the energetics this low in the atmosphere and is not included in
the plot.}
\label{fig:photoheat}
\end{figure}

Our flare simulations exhibit strongly enhanced optical continuum
emission. Figure~\ref{fig:cspec} shows
optical continuum spectra for the preflare atmosphere and a
few times during the F10 (panel a) and F11 (panel b) flares. A
large Balmer jump and noticeable Paschen jump are clearly present
in these spectra. The elevated flux peaking at about 6500 \AA\ is
primarily the result of blackbody emission and indicates heating
has occurred in the photosphere.  The electron beam is unable to
penetrate deeply enough to directly heat the photosphere; instead
the heating in this region is primarily due to radiative
backwarming from the Balmer and Paschen continua. This is shown in
Figure~\ref{fig:photoheat} which plots the energy deposited in
the lower atmosphere as a function of height above the
photosphere. Important contributors to the gain in internal
energy are plotted in Figure~\ref{fig:photoheat}a showing that
energy deposited by radiation is the dominant form of flare
heating in the photosphere. Figure~\ref{fig:photoheat}b plots the
most important contributors to the radiative heating and
indicates that Balmer and Paschen continua significantly heat the
lower atmosphere. The photosphere reaches a maximum temperature
increase of 400~K for F10 and 1200~K for F11.  This
together with the elevated Paschen continuum produces a 32\% and a
129\% increase in the optical continuum (measured at 6000 \AA) for
F10 and F11 respectively.

\section{Comparisons to the Solar Flare Models of A05}
\begin{figure}
\plottwo{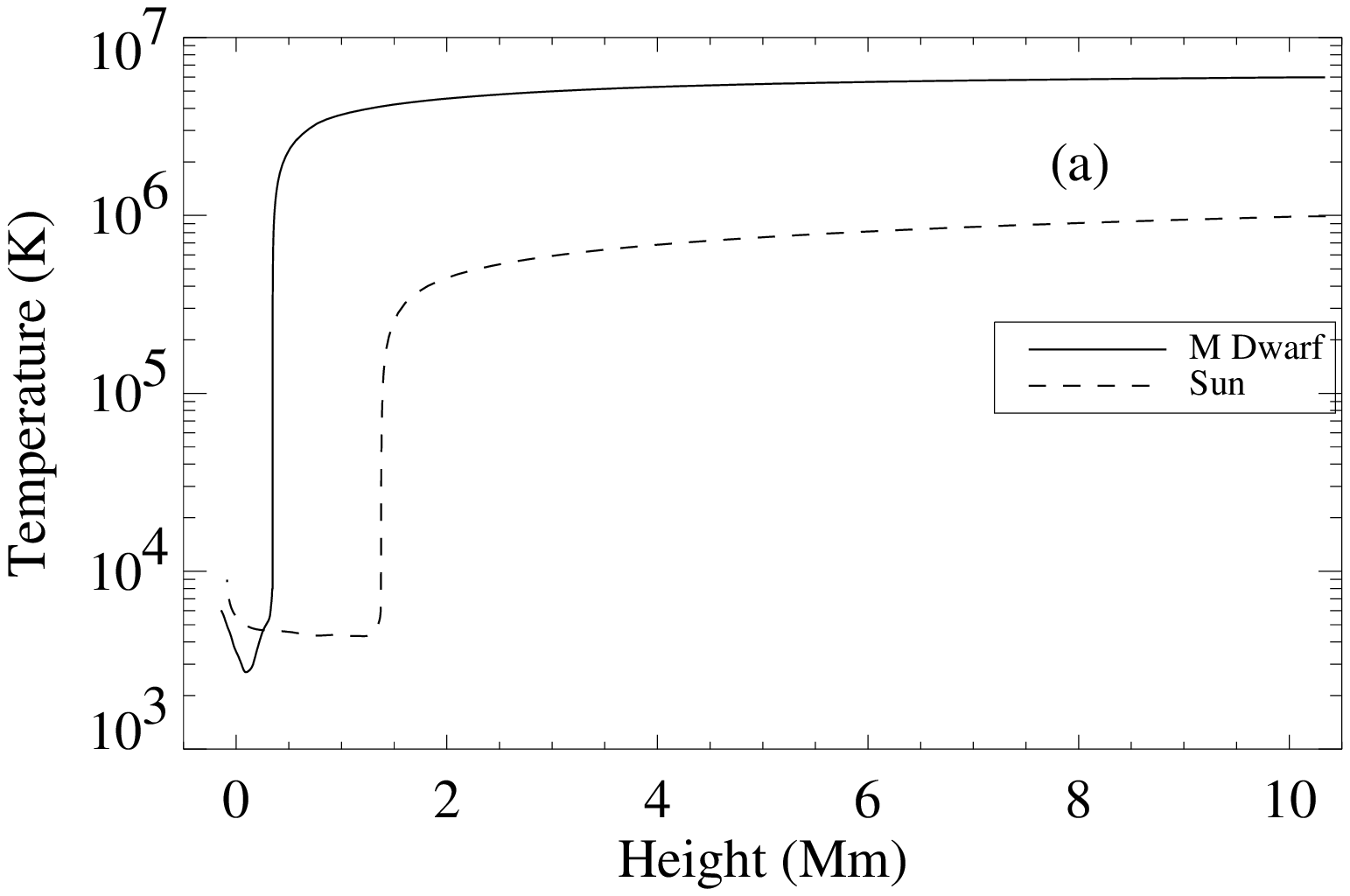}{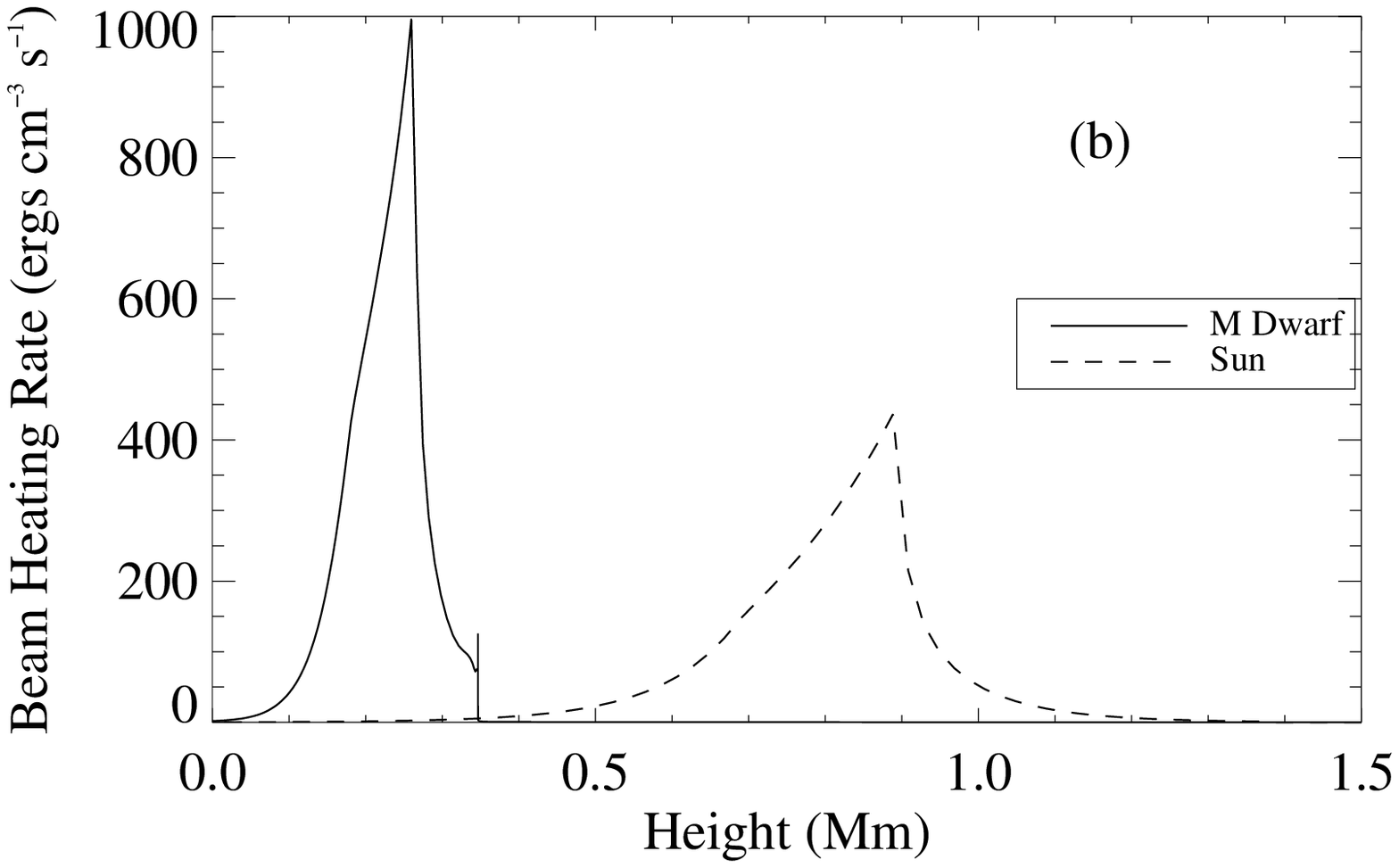}
\caption{(a) Temperature structure of the preflare M dwarf (solid
line) and solar (dashed line) atmospheres.  (b) Initial beam energy
deposition rate for the M dwarf (solid line) and solar (dashed
line) cases.}
\label{fig:pfas}
\end{figure}

Compared to the Sun, an active M dwarf has a higher surface
gravity, a cooler photosphere, and a hotter corona.
Figure~\ref{fig:pfas}a compares the temperature structure of our
M dwarf preflare atmosphere with the solar preflare atmosphere
used in A05. The higher surface gravity on the M dwarf compresses
the atmosphere relative to the Sun, causing the chromosphere and
transition region to be closer in height to the photosphere than
in the solar model. The density of the M dwarf corona is
approximately an order of magnitude greater than the solar
corona.  The compressed atmosphere results in a much narrower
range of beam energy deposition (shown in
Figure~\ref{fig:pfas}b).

Many of the features predicted by the solar models of A05 are
similar to those in the M dwarf simulations presented here. For
example, despite differences in atmospheric structure, both types
of simulations predict chromospheric evaporation waves of similar
speeds. The reason for this is that the beam energy is deposited
in each atmosphere at nearly the same column mass.  Since the same
beam energy is assumed in the solar F10 and M dwarf F10
simulations, and the mass above the beam impact location is nearly
the same, similar velocities should be expected. Another important
similarity is that both types of simulations exhibit an initial
gentle phase followed by explosive eruption in the transition
region when the balance between flare heating and radiative
cooling is broken. However, there is significant difference
between the M dwarf and solar models in the onset time of the
explosive phase. The explosive phase starts at 98~s and
2.5~s for the M dwarf F10 and F11 flares and at 73~s and 1.0~s for
the solar F10 and F11 flares. The reason for this is that since
the M dwarf model has a higher coronal density, it takes longer
for the helium to be fully ionized so that the beam heating
overcomes the radiative cooling.

\begin{figure}
\epsscale{.6}
\plotone{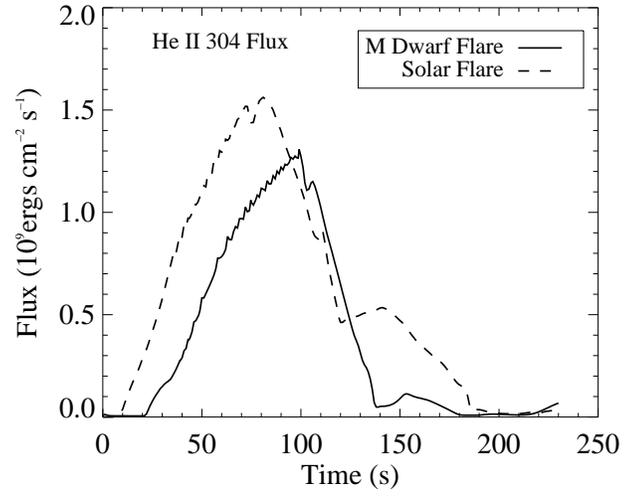}
\caption{Light curves of the \heiilya\ line are plotted for the F10
M dwarf simulation presented in this paper and the F10 solar model
of A05.  The explosive phase begins earlier and there is more total
emission in the solar case.}
\label{fig:heiias}
\end{figure}

\begin{figure}
\epsscale{.6}
\plotone{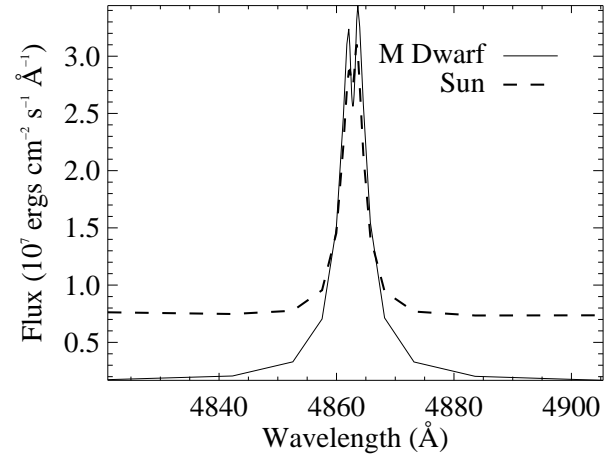}
\caption{\hb\ line profiles at 100~s for the F10 M dwarf simulation
and the F10 model of A05. The line profiles are similar in shape
and maximum intensity, but since the level of the continuum is
higher in the solar case there is less total line emission.}
\label{fig:hbas}
\end{figure}

Another difference is the intensity of the emitted radiation. In
the solar model, the helium-emitting plateau in the transition
region becomes much more extended than in the M dwarf models
(compare the first panels of Fig.~\ref{fig:f10exp} and Fig.~4 of
A05) causing a larger \heiilya\ enhancement in the solar case.
This is seen in Figure~\ref{fig:heiias} which shows light curves
for the \heiilya\ line during the F10 solar and M dwarf
simulations.  In both cases, the \heiilya\ flux peaks at the onset
of the explosive phase.  In contrast, the hydrogen Balmer
line emission is stronger in the M dwarf flare spectrum.  Since the
Sun has a hotter photosphere, the continuum level is much higher
for the solar Balmer lines. The Balmer line profiles for the solar
and M dwarf cases are similar, but since the line to continuum
ratio is higher in the M dwarf case, it has more Balmer line
emission. The \hb\ line profile in Figure~\ref{fig:hbas}
illustrates this effect.

\section{Comparison to M dwarf Flare Observations}
\subsection{Continuum}
\begin{figure}
\epsscale{.6}
\plotone{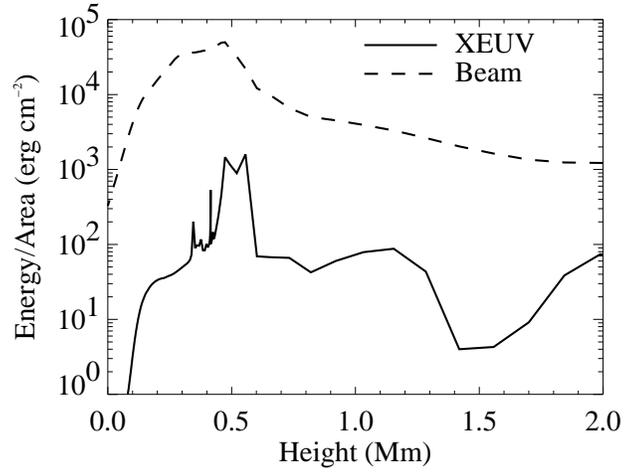}
\caption{Energy deposited during the F10 flare through XEUV
backwarming and electron beam heating is plotted as a function of
atmospheric depth. The XEUV heating accounts for about 1\% of the
total energy deposited.}
\label{fig:xeuv}
\end{figure}

Numerous photometric observations of flares on active M dwarfs have
shown that the continuum spectrum fits well to a blackbody of
approximately 9000 K \citetext{\citealp{hf92, e92, a94}; H03}. This
is puzzling because the electron beam, which is the assumed source
of flare heating, is unable to deposit sufficient
energy in the photosphere to heat it to so high a temperature.
One explanation, proposed by \citet{hf92}, suggested that XEUV
backwarming may be responsible for heating the photosphere during
flares. To test this, we included XEUV backwarming as described in
\S~2. Figure~\ref{fig:xeuv} plots the energy
deposited during the F10 flare due to XEUV emission and compares it
to beam heating. We find that the XEUV heating is about a hundred
times smaller than the beam heating and does not significantly
contribute to heating the photosphere.  As discussed in
\S~\ref{sec:cont} Balmer and Paschen continuum backwarming does
heat the photosphere, but even this is insufficient to raise the
temperature to 9000 K.

\begin{figure}
\epsscale{.5}
\plotone{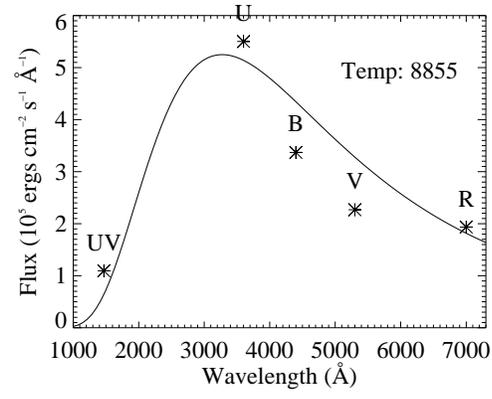}
\caption{The flare spectrum convolved with the \ubvr\ filters and
the corresponding blackbody fit are plotted.  The \emph{UV} point
is obtained by averaging the spectrum over three 30~\AA\ regions
centered at 1469~\AA. The spectrum is best fit with a temperature
of 8855~K.}
\label{fig:bbfit}
\end{figure}

Another explanation is that the observed \ubvr\ enhanced spectrum
may not actually be the result of blackbody emission, but rather
may be due to increased Balmer emission from the chromosphere.
This explanation is supported by these simulations. Despite a lack
of direct photospheric heating, when convolved with the \ubvr\
bandpasses these simulations do produce a spectrum with the general
shape of a blackbody with a temperature of 8000 -- 9000~K.
Figure~\ref{fig:bbfit} shows an example of the convolved
spectrum and blackbody fit. In a fashion similar to H03 we include
an additional measurement in the UV, centered at 1469~\AA. The
spectrum is similar to observed continuum enhancements seen during
flares (cf. Figure~10 of H03), and therefore, it is plausible that
the ubiquitous white light continuum may not be thermal in nature,
but actually a result of Balmer and Paschen emission. However, the
increased \emph{U} band seen in our simulations is the result of a
strongly elevated Balmer jump which has not been observed in
stellar flare spectra \citep{hp91,e92}.

\subsection{Continuum Dimming}
\begin{figure}
\plotone{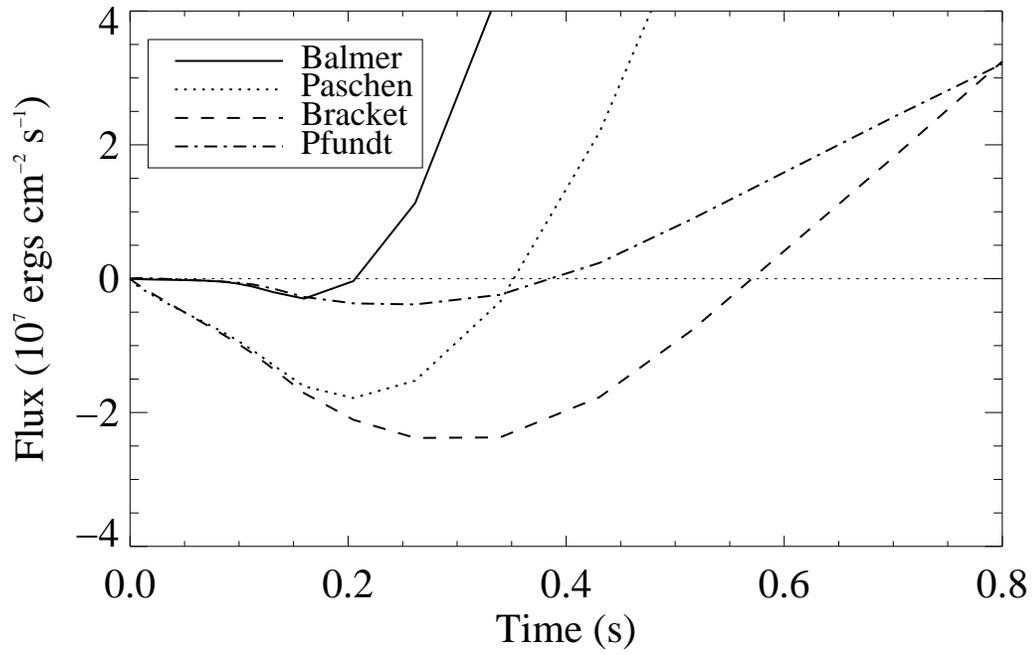}
\caption{Fluxes for the Balmer, Paschen, Bracket
and Pfundt continua are plotted as a function of time. This
plot illustrates the initial continuum dimming seen in the higher
order hydrogen continua.} \label{fig:cdim}
\end{figure}

An initial dimming in the stellar continuum at the onset of a
white light flare has been observed by many authors
\citep{crg80,g82,d88,h95}. In a manner similar to \citet{ah99}
and A05, our simulations also exhibit initial continuum dimming.
Figure~\ref{fig:cdim} shows light curves for
Balmer and higher order hydrogen continua. The continuum dimming
seen here is a result of non-thermal ionization produced by the
electron beam and can be understood as follows. The beam raises
collisional rates in the upper chromosphere causing higher
population densities for the excited states of hydrogen. This
increases the absorption probability for Balmer and higher order
hydrogen continuum photons, which previously would have escaped
and been seen as continuum emission. This results in an initial
decrease in the continuum intensity.  The time between flare
onset and the re-brightening of the continuum is controlled by
the ratio of recombinations to photoionizations. As the flare
evolves, the increased electron density in the upper chromosphere
elevates the recombination rate and begins to reduce the
population excess of the excited states of hydrogen.  This
decreases the photoionization rate, and the continuum
re-brightens.

\subsection{Line Broadening}\label{sec:lbroad}
\begin{figure}
\epsscale{.666}
\plotone{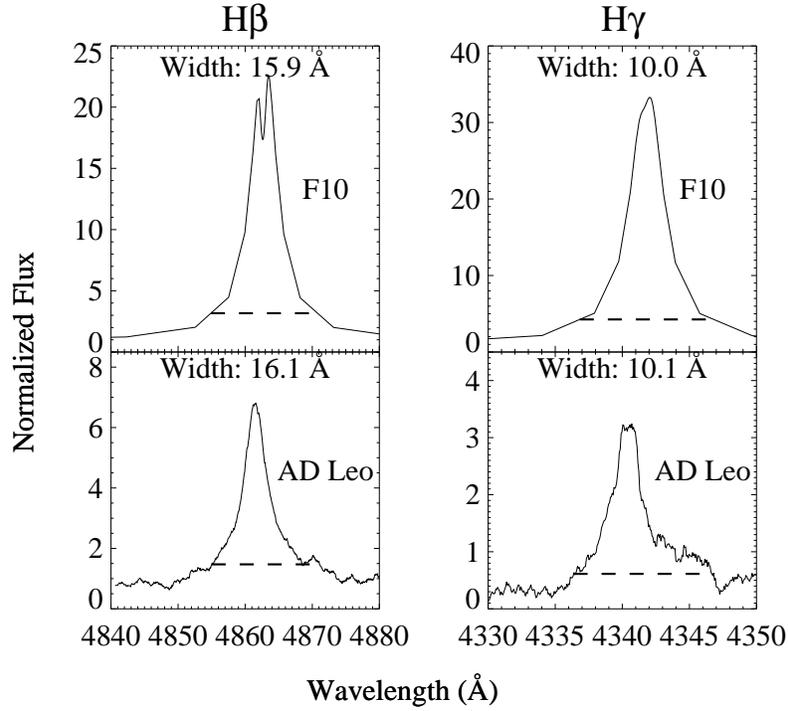}
\caption{Normalized line profiles for \hb\ and \hg\ obtained from
the F10 simulation are compared with observed line profiles at the
peak of Flare 8 in H03. The line widths at 0.1 maximum are
indicated by the dashed lines. The top row shows synthetic profiles
from F10 at a characteristic time during the initial phase (64~s).
The plots on the bottom row show flare spectra (quiescent emission
has been removed) obtained from the peak of flare 8 from H03.}
\label{fig:balwid}
\end{figure}

\begin{figure}
\epsscale{.54}
\plotone{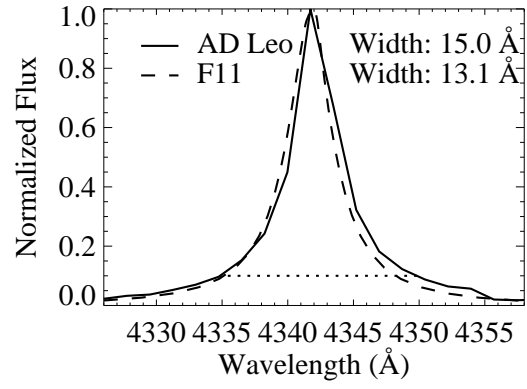}
\caption{F11 \hg\ line profile compared to the
\citet{hp91} AD Leo flare observation.  The
simulated line profile is taken at the time of peak \hg\
emission, and the observed line profile is taken at the peak of the
flare (542~s). In each case, the continuum emission has been
removed and the line profile has been normalized to the line
maximum. The \citet{hp91} line width at 0.1 maximum
is indicated by the dotted line.}
\label{fig:hgbf}
\end{figure}

During the impulsive phase the Balmer lines dramatically increase
in width, as observed by numerous authors \citetext{see for
example, \citealp{d88, 1988MNRAS.235..573P, hp91, e92,
1995A&AS..114..509A}; H03}. Figure~\ref{fig:balwid} shows line
profiles computed during F10 for \hb, and \hg\ and compares them
to line profiles observed during the impulsive peak of Flare 8 of
H03. Similarly, Figure~\ref{fig:hgbf} shows the \hg\ line compared
to the line profile obtained by \citet{hp91}.  In each case the
quiescent line emission has been removed from the line profile. To
aid in measuring the line width, smoothing has been applied to the
H03 observed line profile.  To include the broadened wings, we
measure line widths at one tenth maximum. In both the simulation
and the H03 flare observation, \hb\ increases to a width of about
16 \AA, and \hg\ reaches a peak line width of 10 \AA. The
\citet{hp91} flare was much larger than the H03 flare, and in this
flare \hg\ had a peak width of 15 \AA\ compared to 13.1 \AA\
obtained in our F11 simulation.

The observed line broadening is likely caused by either Stark
broadening from the increased electron density in the flaring
atmosphere, or from turbulent and mass motion flows in the
flaring atmosphere. \citet{d88, e92, 1995A&AS..114..509A} found
that mass flow velocities of $\sim$100 -- 300 km s$^{-1}$ would
be required to produce the observed Balmer line widths. In both
our F10 and F11 models, chromospheric velocities are much less
than these (\hg\ has a maximum velocity of $\sim$10 km s$^{-1}$),
and we find that in these simulations the line broadening is
almost entirely due to the Stark effect.

\subsection{Balmer Decrement}
The Balmer decrement represents the relative energy emitted in
various Balmer lines.  It is defined as the ratio of the excess
flare emission in the Balmer lines relative to a fiducial, often
taken to be \hb\ or \hg. Information about the electron density in
the Balmer line forming region of the atmosphere may be inferred,
with steeper decrements implying smaller electron densities
\citep{2002A&A...383..548G}. A simple explanation for this is
that as densities increase, the lower order lines become thicker
and more energy can escape in the higher order lines, making the
decrement shallower.  Since the decrement is a relative
measure, it makes comparing flares of different sizes
straightforward.  In Table~\ref{table:baldec}, we present Balmer
decrements calculated relative to \hg\ for F10 and F11 and compare
them to several observations.  We have also included the \caiik\
decrement.  For the F10 and F11 simulations we employed a six
level hydrogen atom, and therefore we could not directly calculate
Balmer lines higher than \hg\ in detail.  To obtain line fluxes
for higher order lines we passed a ``snapshot'' of the atmosphere
at the peak of Balmer emission to the static radiative transfer
code, MULTI \citep{c86}. In MULTI we used a 13 level hydrogen
atom, allowing us to model the Balmer lines through H12. Comparing
F11 to F10 shows that for a larger flare, \hg\ becomes relatively
stronger than the other Balmer lines.  All the observed Balmer
decrements more closely match F11 than F10, as expected since
these flares were all relatively large. The decrements for a flare
on AD Leo reported by \citet{hp91} are especially close to F11,
indicating that the flare chromospheric structure likely resembles
the structure predicted by this simulation.

\subsection{Line to Continuum Ratios}
The continuum carries 79\% of the total emission in the F10 flare.
The lines are relatively stronger in the F11 flare with the
continuum carrying only 72\% of the energy. Both model predictions
are considerably smaller than ratios typically observed for M
dwarf flares.  H03 observed a continuum emission to total emission
ratio of $\sim$92\% for five moderate flares on AD Leo.
\citet{1984iue..conf..247R} also observed a ratio of 92\% for a
flare on AD Leo, and for a very large flare \citet{hp91} obtained
a ratio of 86\%. Our over-estimation of the relative line strength
is likely due to the fact that these simulations do not reproduce
the featureless $\sim$9000 K white light continuum observed in M
dwarf flares.  If we add a 9000~K blackbody spectrum to our model
spectrum, we do obtain a line to continuum ratio of more than 90\%
as expected.

Another possible reason for the significant line emission from
our simulations is the shape of the beam energy spectrum.  If we
are over-estimating the concentration of beam energy that
is deposited in the upper chromosphere, the excess heating
will produce too much Balmer emission. The majority of the beam
energy is deposited within a relatively narrow region of the
chromosphere and causes large increases in the Balmer emission.
\citet{1997ApJ...483..496M} have shown, using a Fokker-Planck
description of the electron beam, that the beam energy deposition
is over a wider region and higher in the atmosphere than is
predicted by the analytic expression employed in these
simulations. Less concentrated heating will likely produce
less dramatic Balmer emission and a line to continuum ratio which
more closely matches observations.

\subsection{UV Line Velocities \label{sec:uvlvel}}
\begin{figure}
\epsscale{.8}
\plotone{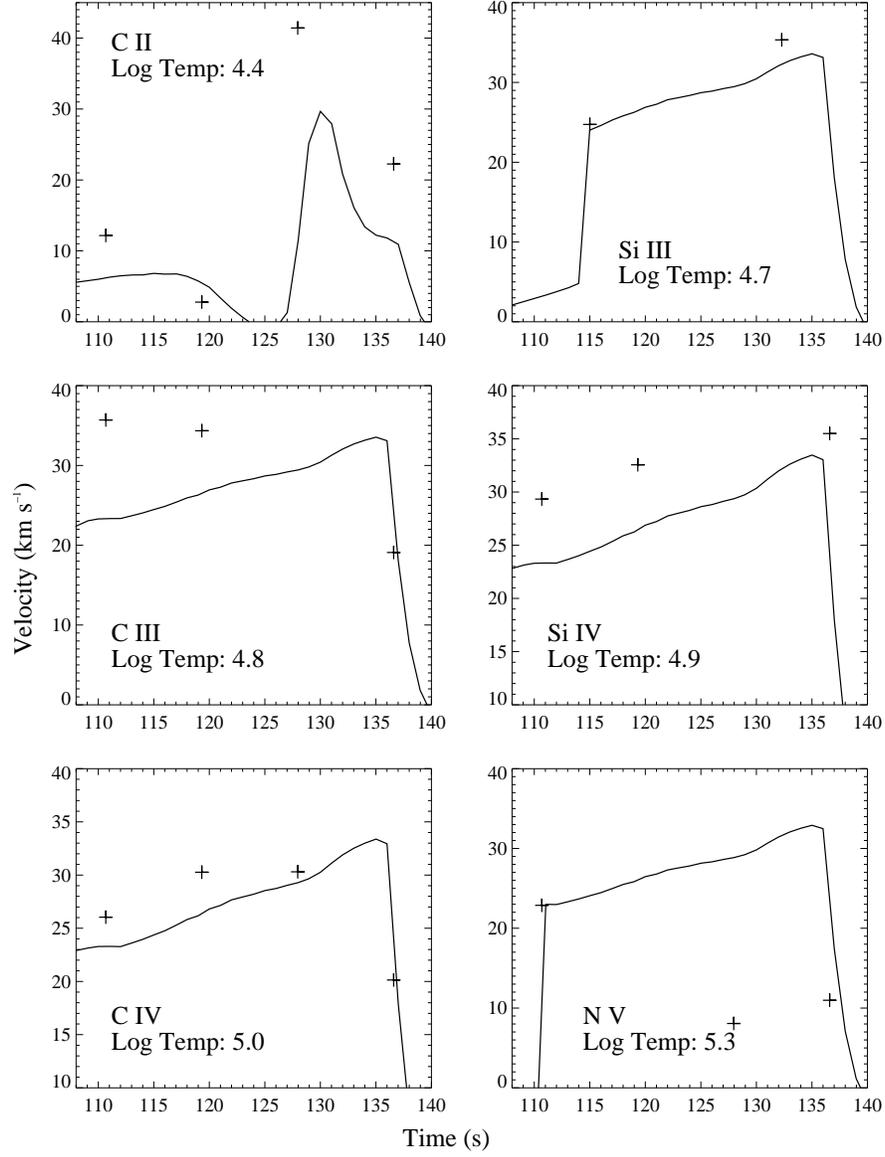}
\caption{F10 plasma velocities are plotted and compared to
velocities obtained during Flare 8 of H03.  The solid lines
plot velocities of the plasma at the temperature
indicated, and the crosses mark the observed velocities. The
labels in each panel indicate the ion measured and its formation
temperature.  The time axis of Flare 8 has been shifted by
7.0~s.}
\label{fig:vel}
\end{figure}

H03 observed numerous transition region and coronal lines with the
HST/STIS instrument during flares on AD Leo. They found that many
of the lines formed in the transition region were red-shifted
during the impulsive phase of the flare, and interpreted this as an
example of chromospheric condensation. To compare our simulations
to these observations, we used the CHIANTI \citep{d97,y03}
database to produce synthetic line profiles for the lines observed
by H03. In our flare simulations, we find multiple regions of the
atmosphere with typical transition region temperatures
($\sim$10$^5$). For example, at 150 s in F10 (panel 3 of
Fig.~\ref{fig:f10exp}) there are three regions with this
temperature  ($z=0.5, 3.1$ and 3.6 Mm). The outer two ($z=3.1$ and
3.6 Mm) are traveling upward producing blue-shifted emission and
the inner one ($z=0.5$ Mm) is moving deeper into the atmosphere
causing red-shifted emission.
In Figure~\ref{fig:vel} we compare the velocity of the downward
directed plasma with the velocities obtained during Flare 8 of
H03. Both the simulation and observations exhibit velocities of 20
-- 40 km s$^{-1}$ in agreement with the condensation velocities
predicted by \citet{1989ApJ...346.1019F}.

\section{Conclusions}
We have constructed radiative hydrodynamic simulations of a
flaring M dwarf atmosphere and used them to explore the
flare-induced optical and UV emission.  We find that the dynamics
of the impulsive flare naturally divide into two phases, an
initial gentle phase followed by a period of explosive increases
in temperature and pressure.  The explosive wave front creates a
high temperature bubble which expands until the wave passes
through the boundary of the flux tube.  The atmosphere then
attains a steady state with a corona and transition region
that are denser than the preflare atmosphere.

Line emission is greatly enhanced in both the moderate (F10) and
strong (F11) flare simulations.  As chromospheric material is
evaporated into the transition region, the \heiilya\ line is
especially enhanced, becoming the strongest line in the spectrum.
Large hydrodynamic waves produced by the flare heating cause
significant Doppler shifting of the line profiles.  The \heiilya\
line exhibits a
peak upward velocity of 129 km~s$^{-1}$, and the \caiik\ line has
a peak downward velocity of 30 km~s$^{-1}$.  We compared the
predicted velocity of our downward-directed condensation wave to
velocities measured in transition region lines by H03, and found
that both models and observations exhibited velocities between 20
-- 40 km~s$^{-1}$.

Stark broadening caused by the increased electron density in the
flaring atmosphere makes the Balmer lines very wide.  We compared
our F10 Balmer profiles to the observations in H03, and found
that for both observations and simulations \hb\ had a width of 16
\AA, and \hg\ had a width of 10 \AA.  We also found that for the
stronger \cite{hp91} flare, the \hg\ line was similar in width
and shaper to the F11 \hg\ line profile.

Our simulations predict elevated optical continuum emission.
A large Balmer jump in the flare spectrum radiatively backwarms
the atmosphere below the temperature minimum region and heats the
photosphere by about 400~K in the F10 case and 1200~K for F11.
With the elevated Paschen continuum this produces a 32\% increase
in the continuum level for F10 and a 129\% increase for F11.  The
predicted photospheric temperature increase is not large enough to
explain the observed blackbody continuum of $\sim$9000 K.  Our
predicted spectrum, when convolved with the \ubvr\ filters, does
produce a spectrum which looks like a 9000 K blackbody, but this
depends on the presence of a large Balmer jump which has not been
seen in M dwarf flares.  The optical continuum exhibits an initial
dimming which also has been observed in M dwarf flares.  In our
simulations, the dimming is a result of non-thermal ionizations due
to the electron beam causing an over-density of excited states of
hydrogen.

We have successfully modeled velocities, line broadening, and
Balmer decrements.  While broad band colors are matched by our
spectrum convolved with filter curves, the continuum does not
match in details, predicting a large Balmer jump and a
noticeable Paschen jump which are not observed.  The formation
of the strong, featureless white light continuum during stellar
flares, therefore, remains a mystery.

Although we have taken great care to make these simulations as
accurate as possible, we have found it necessary to make several
limiting assumptions.  The atmosphere is assumed to be
one-dimensional, plane-parallel, and aligned with the magnetic
field.  This neglects the possibility that radiation can escape
through the sides of the flux tube. The ion and electrons
temperatures are assumed to be the same.  We have also assumed
complete redistribution holds for all transitions. By neglecting
MHD effects we have implicitly assumed that that magnetic pressure
is much higher than the gas pressure and effectively confines the
plasma to the flux tube.

We plan to make several improvements to these models.  As
mentioned above the analytic treatment of \citet{e78} results in
the beam energy being deposited in a relatively narrow region of
the atmosphere causing steep electron density gradients and much
stronger than observed Balmer emission.  We plan to
incorporate a model of the electron beam using Fokker-Planck
kinetic theory.  Preliminary analysis shows a wider range of
energy deposition with a larger portion of the beam deposited in
higher regions of the atmosphere than predicted by \citet{e78}.
This will also provide more direct heating to the corona.  With
these changes, we hope to be able to model a larger flare loop,
as is probably more appropriate for the stellar flare case.
This may allow the stellar flare explosive wave to pass
through the top of the loop, and thereby allow the flare loop to
obtain a steady state as in the solar case.
\acknowledgments
This work has been partially funded by NSF grant AST02-05875 and
HST grants AR-10312 and GO-8613. The computations presented here
were carried out on the Astronomy Condor Network at the
University of Washington, and we would especially like to thank
John Bochanski for contributing many hours of computer time.

\clearpage
\clearpage
\clearpage
\clearpage
\clearpage
\clearpage
\clearpage
\clearpage
\clearpage
\clearpage
\clearpage
\clearpage
\clearpage
\clearpage
\clearpage

\clearpage
\begin{deluxetable}{lcccccc}
\tablewidth{0pt}
\tablecaption{Balmer Decrement \label{table:baldec}}
\tablecolumns{7}
\tablehead{ \colhead{Source} & \colhead{\hb} &  \colhead{\hg} &
\colhead{H$\delta$} & \colhead{H8} & \colhead{H9} &
\colhead{\caiik} }
\startdata
F10 Flare & 1.28 & 1.00 & 0.96 & 0.82 & 0.74 & 0.96\\
F11 Flare & 1.19 & 1.00 & 0.89 & 0.64 & 0.53 & 0.12\\
AD Leo (HP91)\tablenotemark{a} & 1.24 & 1.00 & 0.85 & 0.64 & 0.47 & 0.22\\
AD Leo (R84)\tablenotemark{b} & \nodata & 1.00 & 0.85 & 0.53 & 0.39 & 0.15 \\
UV Cet (P88)\tablenotemark{c} & \nodata & 1.00 & 0.69 & 0.47 & 0.42 & 0.83 \\
UV Cet (E92)\tablenotemark{d} & \nodata & 1.00 & 0.96 & 0.46 & 0.34 & 0.09 \\
YZ CMi\tablenotemark{e} & \nodata & 1.00 & 0.69 & 0.54 & 0.45 & \nodata \\
EZ Aqr\tablenotemark{f} & 1.38 & 1.00 & 0.83 & 0.54 & 0.36 &
\nodata \\
Solar M-class Flare\tablenotemark{g} & 1.74 & 1.00 & 0.80 & 0.63 &
\nodata & \nodata
\enddata
\tablenotetext{a}{Impulsive phase. \citep{hp91}}
\tablenotetext{b}{Impulsive phase. \citep{1984iue..conf..247R}}
\tablenotetext{c}{Flare maximum. \citep{1988MNRAS.235..573P}}
\tablenotetext{d}{Impulsive phase. \citep{e92}}
\tablenotetext{e}{Flare average. \citep{d88}}
\tablenotetext{f}{Flare maximum of Flare E. (Jevremovic et
al. 1998)} 
\tablenotetext{g}{Flare maximum. \citep{1997ApJS..112..221J}}
\end{deluxetable}

\end{document}